\documentclass{nolta}
\usepackage[dvips]{graphicx}
\usepackage{bm,amsmath,amssymb,subfigure}

\Vol{2}
\No{3}
\Year{2011}
\Month{7}
\title{A criterion for timescale decomposition of external inputs for generalized phase reduction of limit-cycle oscillators}

\AUTHOR{%
\author[a]{Wataru Kurebayashi}{1}, 
\author{Sho Shirasaka}{1},
\author{Hiroya Nakao}{1}
}

\AFFILIATE{%
\affiliate{Graduate School of Information Science and Engineering,
	Tokyo Institute of Technology --- 2-12-1 O-okayama, Meguro-ku, Tokyo 152-8552, Japan}{1}
}

\EMAIL{%
\email{kurebayashi.w.aa@m.titech.ac.jp}{a}
}

\usepackage{color}
\usepackage{ulem}
\newcommand{\INS}[1]{#1}

\received{10}{27}{20XX}
\revised{12}{29}{20XX}
\published{7}{1}{20XX}

\begin{document}

\begin{abstract} 
The phase reduction method is a dimension reduction method for weakly driven limit-cycle oscillators,
which has played an important role in the theoretical analysis of synchronization phenomena.
Recently, we proposed a generalization of the phase reduction method [W. Kurebayashi {\it et al}., {\it Phys. Rev. Lett.} {\bf 111}, 2013].
This generalized phase reduction method can robustly predict the dynamics of strongly driven oscillators, for which the conventional phase reduction method fails.
In this generalized method, the external input to the oscillator should be properly decomposed into \INS{a slowly varying component and remaining weak fluctuations}.
In this paper, we propose a simple criterion for timescale decomposition of the external input,
which gives accurate prediction of the phase dynamics and enables us to systematically apply the generalized phase reduction method to a general class of limit-cycle oscillators.
The validity of the criterion is confirmed by numerical simulations.
\end{abstract}


\maketitle 

\section{Introduction}

Synchronization of limit-cycle oscillators is a ubiquitous phenomenon that has been widely studied in many disciplines including
physics, chemistry, biology, mechanical engineering, and electrical engineering~\cite{kuramoto,winfree,pikovsky}.
The phase reduction method~\cite{kuramoto} has played a key role
in the theoretical analysis of synchronization phenomena of weakly driven limit-cycle oscillators.
This method enables us to reduce the dynamical equation of a high-dimensional limit-cycle oscillator to a one-dimensional phase equation,
which facilitates theoretical analysis of various synchronization phenomena.
Recently, engineering applications of the phase reduction method, {\it e.g.},
\INS{dynamical analysis} and optimal design of circuit and other oscillators~\INS{\cite{Demir2000,Tanaka2002,Nagashima2014,Vytyaz2009,Maffezzoni2010a,Kure2014}}
and optimal control of periodically spiking neurons~\cite{Dasanayake2011,Nabi2012,Moehlis2014}, have been actively studied.

However, the conventional phase reduction method has a drawback in practical applications,
{\it i.e.}, it works only when the external input to the oscillator can be assumed sufficiently weak.
This limitation significantly narrows the applicability of the conventional method,
because the weakness of the input cannot be assumed in many practical applications.
In order to overcome this limitation, we recently proposed a generalized phase reduction method~\cite{Kurebayashi2013},
which robustly works even for largely varying inputs
under appropriate conditions.
Using the generalized method, we can theoretically analyze the phase dynamics of strongly driven limit-cycle oscillators.

When we use the generalized phase reduction method, we need to decompose the external input to the oscillator
into a slowly varying low-frequency component and sufficiently weak fluctuations.
Though this timescale decomposition can significantly affect the accuracy of the resulting phase equation,
Ref.~\cite{Kurebayashi2013} did not provide an explicit criterion for decomposing the external input.
In the present study, we propose a \INS{simple practical} criterion to decompose the external input into low-frequency and high-frequency components, which \INS{yields a reasonable approximation of} the oscillator dynamics.
The proposed decomposition method will enable us to systematically apply the generalized phase reduction method to the analysis of various synchronization phenomena.

\section{Generalized phase reduction method}

We consider a limit-cycle oscillator driven by a general input $\bm{I}(t) = [I_1(t),\ldots,I_m(t)]^{\top} \in \mathbb{R}^m$ \INS{that smoothly depends on time $t$}, described by
\begin{align}
 \frac{d\bm{X}(t)}{dt} &= \bm{F}(\bm{X}(t),\bm{I}(t)), \label{eq. lco}
\end{align}
where $\bm{X}=[X_1,\ldots,X_n]^{\top}\in\mathbb{R}^n$ is the state of the oscillator
and $\bm{F}(\bm{X},\bm{I})=[F_1(\bm{X},\bm{I}),\ldots,F_n(\bm{X},\bm{I})]^{\top}$ $\in\mathbb{R}^n$ is a vector field
that represents the dynamics of the oscillator.
We assume that there exists a \INS{finite} interval $A \subset \mathbb{R}^m$ of the input value $\bm{I}$ such that the vector field $\bm{F}(\bm{X},\bm{I})$
has a stable periodic orbit $\bm{X}_0(t,\bm{I})$ with period $T(\bm{I})$ and
frequency $\omega(\bm{I}):=2\pi/T(\bm{I})$ when the input $\bm{I}$ is kept constant, and that this periodic orbit smoothly depends on $\bm{I} \in A$.

Using the generalized phase reduction method~\cite{Kurebayashi2013}, we can reduce the high-dimensional dynamics of the limit-cycle oscillator
described by Eq.~(\ref{eq. lco}) to a one-dimensional phase equation.
We define a generalized asymptotic phase $\Theta(\bm{X},\bm{I})$ of the limit cycle $\bm{X}_0(t,\bm{I})$ that satisfies
\begin{align}
 \frac{\partial \Theta(\bm{X},\bm{I})}{\partial \bm{X}} \cdot \bm{F}(\bm{X},\bm{I}) &= \omega(\bm{I}).
\end{align}
for each constant $\bm{I} \in A$. We then decompose the input $\bm{I}(t)$ into a low-frequency component $\bm{q}(\epsilon t) = [q_1(\epsilon t),\ldots,q_m(\epsilon t)]^{\top} \in A$
and a high-frequency component $\sigma \bm{p}(t)  = \sigma [p_1(t),\ldots,p_m(t)]^{\top} \in \mathbb{R}^m$ as
\begin{align}
 \bm{I}(t) &= \bm{q}(\epsilon t) + \sigma \bm{p}(t), \label{eq. inp decomp}
\end{align}
where the low-frequency component $\bm{q}(\epsilon t)$ is assumed to vary slowly as compared to the amplitude relaxation time of the oscillator,
and \INS{the high-frequency component $\sigma \bm{p}(t)$ is assumed to be sufficiently weak.}
The small parameters $\epsilon$ and $\sigma$ represent the slow timescale of the low-frequency component $\bm{q}(\epsilon t)$
and the intensity of the high-frequency component $\sigma\bm{p}(t)$, respectively.
We introduce a phase variable $\theta(t)$ representing the state of the oscillator as
\begin{align}
 \theta(t) &= \Theta(\bm{X}(t),\bm{q}(\epsilon t)), \label{eq. phase def}
\end{align}
\INS{which depends on the slow low-frequency component $\bm{q}(\epsilon t)$ of the input.}
We also define a conventional phase sensitivity function $\bm{Z}(\theta,\bm{q}) \in \mathbb{R}^n$ and two other sensitivity functions $\bm{\xi}(\theta,\bm{q}) \in \mathbb{R}^m$, $\bm{\zeta}(\theta,\bm{q}) \in \mathbb{R}^m$ as follows:
\begin{align}
 \bm{Z}(\theta,\bm{q}) &= \left. \frac{\partial\Theta(\bm{X},\bm{q})}{\partial\bm{X}} \right|_{\bm{X}=\bm{X}_0(\theta/\omega(\bm{q}),\bm{q})}, \\
 \bm{\xi}(\theta,\bm{q}) &= \left. \frac{\partial\Theta(\bm{X},\bm{q})}{\partial\bm{q}} \right|_{\bm{X}=\bm{X}_0(\theta/\omega(\bm{q}),\bm{q})}, \\
 \bm{\zeta}(\theta,\bm{q}) &= \left. \bm{G}(\bm{X},\bm{q})^{\top} \bm{Z}(\theta,\bm{q}) \right|_{\bm{X}=\bm{X}_0(\theta/\omega(\bm{q}),\bm{q})},
\label{eq:G}
\end{align}
where the $(j,k)$-th element of $\bm{G}(\bm{X},\bm{q}) \in \mathbb{R}^{n \times m}$
is given by $G^{(j,k)}(\bm{X},\bm{q}) := \frac{\partial F_j(\bm{X},\bm{q})}{\partial q_k}$.
Then, the dynamics of the phase variable $\theta(t)$ is described by the following generalized phase equation~\cite{Kurebayashi2013}:
\begin{align}
 \frac{d\theta(t)}{dt} &= \omega(\bm{q}(\epsilon t)) + \epsilon \bm{\xi}(\theta,\bm{q}(\epsilon t)) \cdot \dot{\bm{q}}(\epsilon t)
 + \sigma \bm{\zeta}(\theta,\bm{q}(\epsilon t)) \cdot \bm{p}(t) \cr
& \quad + O\left(\frac{\epsilon^2}{\lambda(\bm{q}(\epsilon t))^2},\frac{\epsilon\sigma}{\lambda(\bm{q}(\epsilon t))},
\frac{\sigma^2}{\lambda(\bm{q}(\epsilon t))},\frac{\epsilon\sigma}{\lambda(\bm{q}(\epsilon t))^2}\right).
 \label{eq. phase eq}
\end{align}
Here, $\dot{\bm{q}}(t)$ denotes $d\bm{q}(\epsilon t) / d(\epsilon t)$ 
and \INS{the function $\lambda(\bm{I})$ is the absolute value of the second largest Floquet exponent of the oscillator (\ref{eq. lco})
driven by a constant input $\bm{I} \in A$}, which characterizes the amplitude relaxation time of the oscillator.
The generalized phase equation (\ref{eq. phase eq}) is valid when 
\begin{align}
 \frac{\sigma}{\lambda(\bm{q}(\epsilon t))} &\ll 1 \quad {\rm and} \quad \frac{\epsilon}{\lambda(\bm{q}(\epsilon t))^2} \ll 1,
\end{align}
\INS{i.e., when the amplitude relaxation is sufficiently fast (see Ref.~\cite{Kurebayashi2013} for a detailed discussion). We hereafter assume that these conditions are satisfied.}

\section{Simple criterion for timescale decomposition of external inputs}

In the generalized phase reduction method, we need to decompose the input $\bm{I}(t)$ as in Eq.~(\ref{eq. inp decomp}).
How to decompose the input $\bm{I}(t)$ is an important problem, which can significantly affect the accuracy of the generalized phase equation (\ref{eq. phase eq}).
Our aim in this paper is to propose a simple criterion for choosing the threshold frequency $\Omega_d$ that gives a \INS{reasonable decomposition} of the input into low-frequency and high-frequency components for approximating the dynamics of the oscillator.
\INS{In our derivation, the essential parameter for the timescale decomposition is the amplitude relaxation time of the oscillator; the statistical property of the external input $\bm{I}(t)$ (e.g., the power spectrum) is not important.}

We define the decomposition of the input ${\bm I}(t)$ by a linear filter $f(\tau)$ as follows:
\begin{align}
 \bm{q}(\epsilon t) &= \int_{-\infty}^{+\infty} \bm{I}(t-\tau) f(\tau) d\tau, \\
 \sigma\bm{p}(t) &= \bm{I}(t) - \bm{q}(\epsilon t),
\end{align}
where $f(\tau)$ is assumed to be an ideal low-pass filter with the cutoff frequency $\Omega_d$,
{\it i.e.}, its amplitude response $A(\Omega):=|\int_{-\infty}^{+\infty}f(\tau)e^{-i\Omega\tau}d\tau|$ of $f(\tau)$ is given by
\begin{align}
 A(\Omega) &= \left\{
							\begin{array}{cl}
							 1 & (|\Omega| < \Omega_d), \\
							 0 & ({\rm otherwise}).
							\end{array}
							\right.
\end{align}

As discussed in \INS{Appendix A}, we can describe the dynamics
of a limit-cycle oscillator by the phase and amplitude variables, where the amplitude variable represents the deviation of the oscillator state from the periodic orbit.
In particular, when the oscillator state $\bm{X}(t)$ is two-dimensional,
it can be fully described by a phase variable $\theta(t)$ defined in Eq.~(\ref{eq. phase def})
and an amplitude variable $r(t)$ defined as
\begin{align}
 r(t) &= R(\bm{X}(t),\bm{q}(\epsilon t)), \label{eq. def r}
\end{align}
where the function $R(\bm{X},\bm{I})$ of $\bm{X} \in \mathbb{R}^n$ and $\bm{I} \in \mathbb{R}^m$ satisfies
\begin{align}
 \frac{\partial R(\bm{X},\bm{I})}{\partial \bm{X}} \cdot \bm{F}(\bm{X},\bm{I}) &= -\lambda(\bm{I})R(\bm{X},\bm{I}).
\end{align}
\INS{As shown in Appendix A, we can derive} the following dynamical equation for the amplitude variable $r(t)$:
\begin{align}
 \frac{dr(t)}{dt} &= -\lambda(\bm{q}(\epsilon t))r +\sigma\bm{\zeta}_r(\theta,r,\bm{q}(\epsilon t)) \cdot \bm{p}(t)
 + \epsilon\bm{\xi}_r(\theta,r,\bm{q}(\epsilon t)) \cdot\dot{\bm{q}}(\epsilon t)+O(\sigma^2), \label{eq. r4 again}
\end{align}
where $\bm{\zeta}_r(\theta,r,\bm{q}) = \bm{G}(\bm{X},\bm{q})^{\top} \frac{\partial R(\bm{X},\bm{q})}{\partial \bm{X}}\big|_{\bm{X}=\tilde{\bm{X}}(\theta,r,\bm{q})}$,
$\bm{\xi}_r(\theta,r,\bm{q}) = \frac{\partial R(\bm{X},\bm{q})}{\partial \bm{q}}\big|_{\bm{X}=\tilde{\bm{X}}(\theta,r,\bm{q})}$,
and $\tilde{\bm{X}}(\theta,r,\bm{q})$ is a state point in $\mathbb{R}^n$ satisfying
$\Theta(\tilde{\bm{X}},\bm{q})=\theta$ and $R(\tilde{\bm{X}},\bm{q})=r$.
\INS{This equation shows that the amplitude $r(t)$ fluctuates around $r=0$ due to the external input.}

The approximation error of the generalized phase equation (\ref{eq. phase eq}) is $O(r)$ \INS{(see Appendix A)}.
Thus, we can minimize the approximation error by minimizing the deviation $r(t)$ from the periodic orbit.
\INS{As shown in Appendix B,} we can approximate $\bm{\xi}_r(\theta,0,\bm{I})$ by $\bm{\zeta}_r(\theta,0,\bm{I})$ as follows:
\begin{align}
 \bm{\xi}_r(\theta,0,\bm{I}) &= \frac{1}{\lambda(\bm{I})} \bm{\zeta}_r(\theta,0,\bm{I}) + O\left(\frac{1}{\lambda(\bm{I})^2}\right).
 \label{eq. der1}
\end{align}
\INS{Moreover, from Eq.~(\ref{eq. r4 again}), the order of $r$ can be evaluated as follows: 
\begin{align}
 r &= O\left(\frac{\epsilon}{\lambda(\bm{q}(\epsilon t))},\frac{\sigma}{\lambda(\bm{q}(\epsilon t))}\right).
 \label{eq. der2}
\end{align}
By plugging Eqs.~(\ref{eq. der1}) and (\ref{eq. der2}) into Eq.~(\ref{eq. r4 again}), we can obtain
\begin{align}
 \frac{dr(t)}{dt} &= -\lambda(\bm{q}(\epsilon t))r +\bm{\zeta}_r(\theta,0,\bm{q}(\epsilon t)) \cdot \tilde{\bm{I}}(t)
 + O\left(\frac{\epsilon^2}{\lambda(\bm{q}(\epsilon t))},
 \frac{\epsilon\sigma}{\lambda(\bm{q}(\epsilon t))},\frac{\sigma^2}{\lambda(\bm{q}(\epsilon t))} \right). \label{eq. r4 again2}
\end{align}
where $\tilde{\bm{I}}(t)$ is a transformed external input, whose $j$-th element is given by
\begin{align}
 \tilde{I}_j(t) := \sigma p_j(t) + \frac{\epsilon}{\lambda(\bm{q}(\epsilon t))}\dot{q}_j(\epsilon t),
 \label{eq:input2}
\end{align}
for $j = 1,\ldots,m$.
We define the variance $V_j(\Omega_d)$ of $\tilde{I}_j(t)$ as
\begin{align}
 V_j(\Omega_d) &= \lim_{\tau\to\infty} \frac{1}{\tau} \int_{0}^{\tau} [\tilde{I}_j(t)]^2 dt,
 \label{eq. var def}
\end{align}
which is finite because ${\bm I}(t)$ is smooth and bounded.
For smaller $V_{j}(\Omega_{d})$, the fluctuation of $r(t)$ becomes smaller and the generalized phase equation will give more precise prediction of the phase dynamics.
}

\INS{
To derive a simple criterion for determining the threshold frequency $\Omega_{d}$, 
we assume that the decay rate $\lambda(\bm{q}(\epsilon t))$ of $r(t)$
does not vary too violently and thus its typical values can be characterized by 
\begin{align}
 \lambda_c &:= \lambda( \overline{\bm q(\epsilon t)} ) = \lambda\left( \lim_{\tau\to\infty} \frac{1}{\tau} \int_{0}^{\tau} \bm{q}(\epsilon t) dt \right),
\end{align}
where $\overline{\bm q(\epsilon t)}$ is the long-time average of the slowly varying part of the input, ${\bm q}(\epsilon t)$.
Though this is a rather rough characterization of the decay rate of $r(t)$, it enables us to derive a simple criterion for the threshold frequency.
Replacing $\lambda(\bm{q}(\epsilon t))$ in Eq.~(\ref{eq:input2}) with $\lambda_c$, Eq.~(\ref{eq. var def}) can be estimated as follows:
\begin{align}
 V_j(\Omega_d) &\approx 2\int_0^{\Omega_d} \frac{\Omega^2}{\lambda_c^2} P_j(\Omega) d\Omega + 2\int_{\Omega_d}^{\infty} P_j(\Omega) d\Omega,
\end{align}
where $P_j(\Omega)$ is the power spectrum of $I_j(t)$.
The optimal threshold frequency $\Omega_d = \Omega^*_d$ that minimizes this approximate variance $V_j(\Omega_d)$ can be determined as
\begin{align}
 \Omega^*_d &= \lambda_c, \label{eq. formula}
\end{align}
because this $\Omega^*_d$ satisfies
\begin{align}
 V_j(\Omega_d) - V_j(\Omega^*_d) &= 2\int_{\Omega_d}^{\Omega^*_d} \left( 1 - \frac{\Omega_d^2}{\lambda_c^2} \right) P_j(\Omega_d) d\Omega \ge 0, \quad {\rm for} \quad \Omega_d \le \Omega^*_d, \\
 V_j(\Omega_d) - V_j(\Omega^*_d) &= 2\int_{\Omega^*_d}^{\Omega_d} \left( \frac{\Omega_d^2}{\lambda_c^2} - 1 \right) P_j(\Omega_d) d\Omega \ge 0, \quad {\rm for} \quad \Omega_d \ge \Omega^*_d.
\end{align}
Thus, under the above approximating assumptions, the optimal timescale for the decomposition of the input that minimizes the variance $V_j(\Omega_{d})$ of the input coincides with the characteristic amplitude relaxation time of the oscillator.}

We propose Eq.~(\ref{eq. formula}) as a simple criterion for choosing the value of the threshold frequency $\Omega_{d}$.  It gives the optimal $\Omega_{d}$
for predicting the oscillator dynamics when $\lambda(\bm{q}(\epsilon t))$ is strictly constant, and is expected to provide \INS{a reasonable prediction even if $\lambda(\bm{q}(\epsilon t))$ varies slowly}.
\INS{The criterion~(\ref{eq. formula}) is valid for general external inputs, including periodic signals with delta-peaked power spectra, because we did not introduce any assumptions on the statistical property of the input in the derivation of the criterion~(\ref{eq. formula}).}
Also, though we assumed that the state of the oscillator is two-dimensional for simplicity, the above result can be generalized to higher-dimensional cases by regarding $\lambda(\bm{q})$ as the absolute value of the second largest Floquet exponent among the $n$ Floquet exponents of the oscillator, because the deviation from the periodic orbit is dominated by the slowest amplitude mode characterized by the second largest Floquet exponent.

\section{Numerical simulations}

\begin{figure}[t]
\begin{center}
 \subfigure[$\Omega_d=0.5$]{\includegraphics[width=7cm]{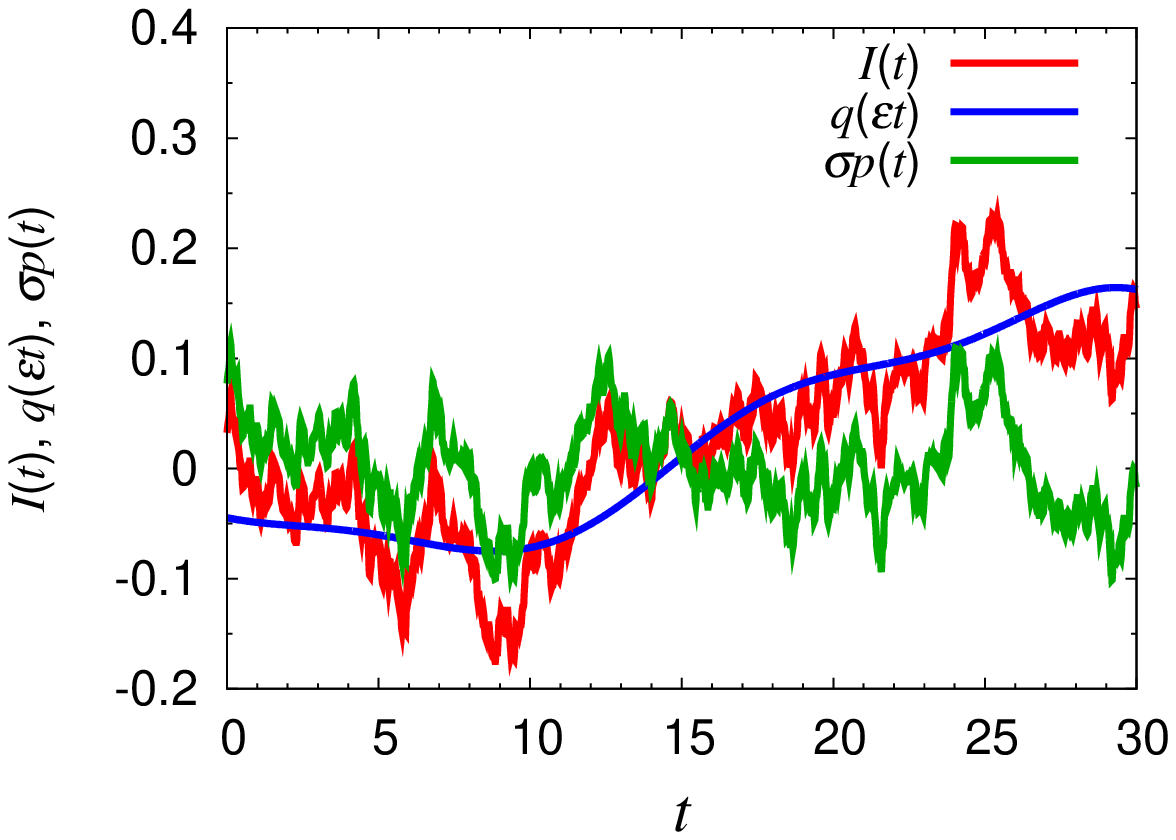}}
 \subfigure[$\Omega_d=2$]{\includegraphics[width=7cm]{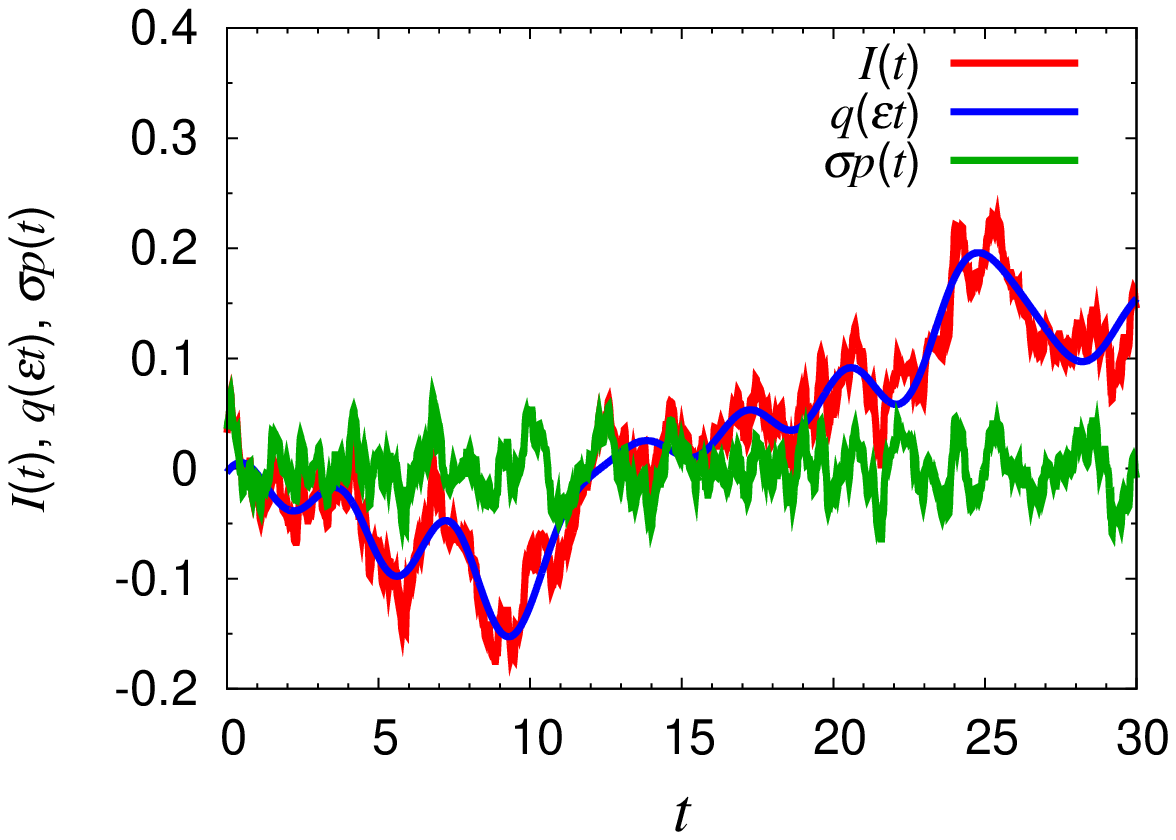}}
 \subfigure[$\Omega_d=5$]{\includegraphics[width=7cm]{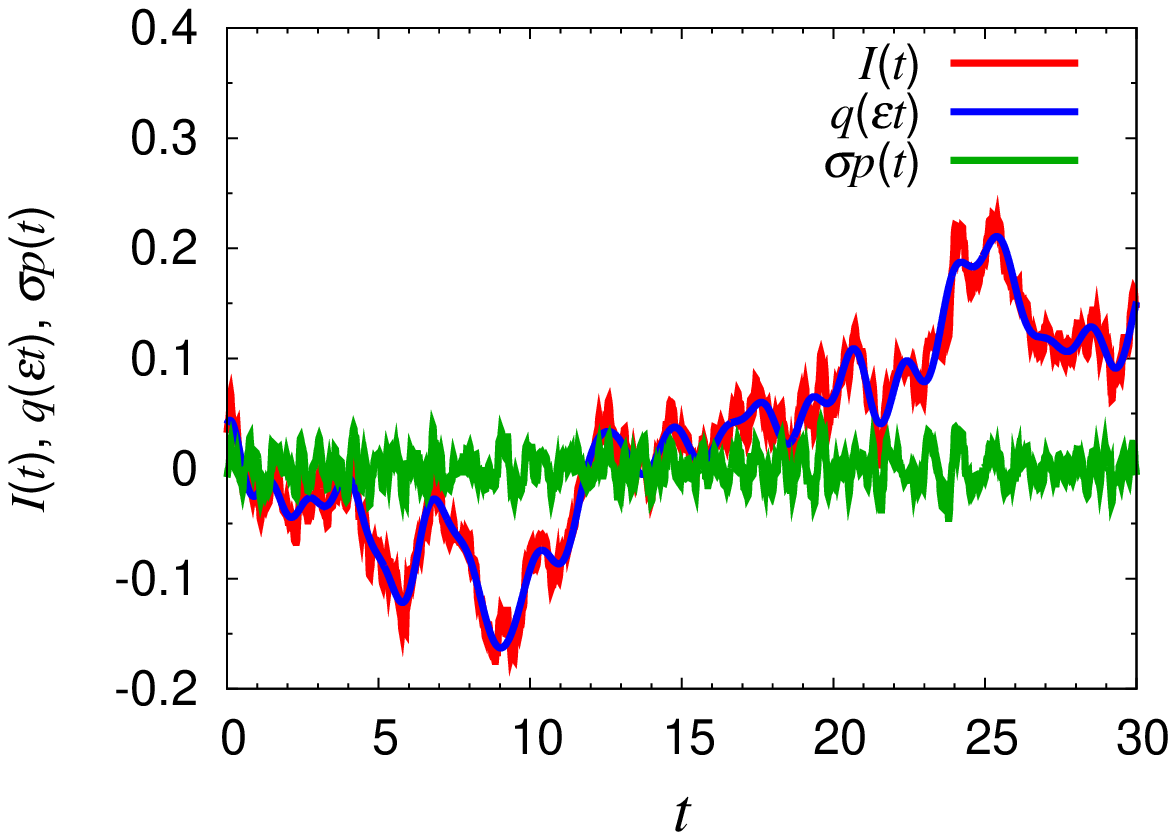}}
 \subfigure[$\Omega_d=10$]{\includegraphics[width=7cm]{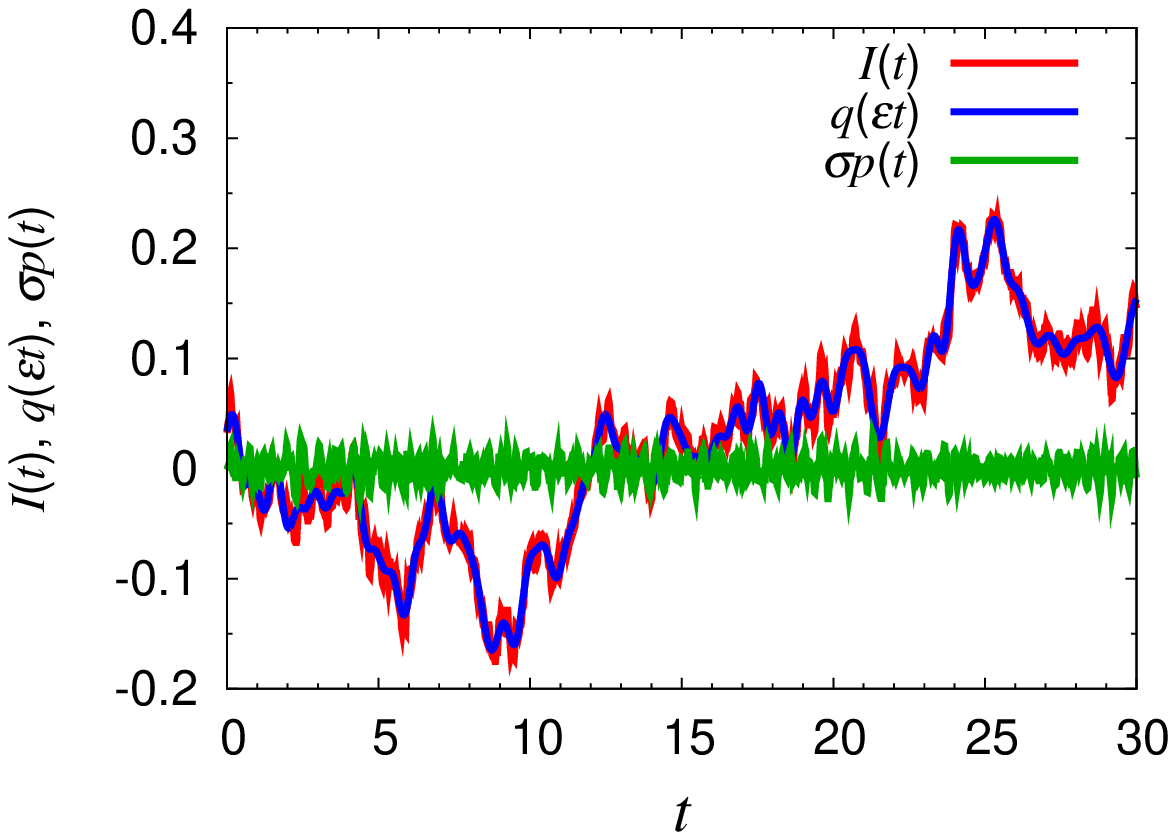}}
\end{center}
\caption{Time series of the input $I(t)$ to the oscillator for $\gamma=5$, and the low-frequency and high-frequency components
 $q(\epsilon t)$ and $\sigma p(t)$ decomposed by low-pass filters with the threshold frequencies $\Omega_d=0.5,2,5$ and $10$.}
 \label{fig. 1}
\end{figure}

\begin{figure}[t]
\begin{center}
\includegraphics[width=9cm]{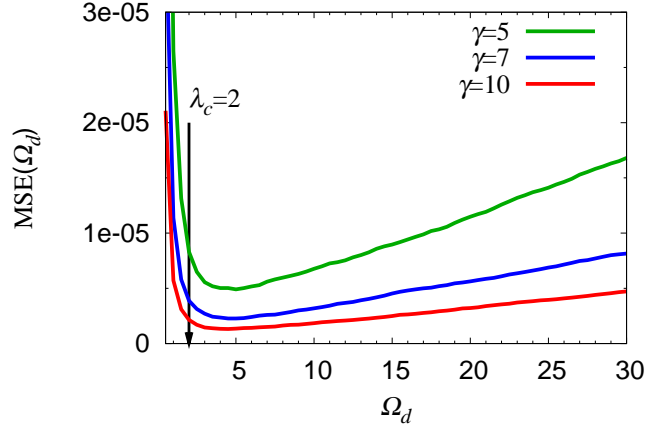}
\end{center}
\caption{Mean square errors of the generalized phase equation (\ref{eq. phase eq})
versus the threshold frequency $\Omega_d$ for $\gamma=5,7$ and $10$.
The arrow represents the threshold frequency $\Omega_d=2$ given by the criterion~(\ref{eq. formula}).}
\label{fig. 2}
\end{figure}

In order to confirm the validity of our criterion~(\ref{eq. formula}), we performed numerical simulations for evaluating
the effect of the threshold frequency $\Omega_d$ on the approximation accuracy of the generalized phase equation (\ref{eq. phase eq}).
In the simulation, we use a modified Stuart-Landau oscillator~\cite{kuramoto} defined as
\begin{align}
 \frac{dx(t)}{dt} &= F_1(x,y,I(t)) := e^{2I(t)} (x-y-I) - [(x-I(t))^2 + y^2](x-I(t)), \label{eq. msl1}\\
 \frac{dy(t)}{dt} &= F_2(x,y,I(t)) := e^{2I(t)} (x+y-I) - [(x-I(t))^2 + y^2]y, \label{eq. msl2}
\end{align}
where $\bm{X}=[x,y]^{\top}$ is a state variable, and $I(t)$ is the input to the oscillator that is
decomposed into a low-frequency component $q(\epsilon t)$ and a high-frequency component $\sigma p(t)$.
For this oscillator, the absolute value of the second largest Floquet exponent is $\lambda(q)=2e^{2q}$.

Defining the phase $\theta(t)$ for this oscillator as
\begin{align}
 \theta(t) &= \Theta(x(t),y(t),q(\epsilon t)) = \tan^{-1}\frac{y(t)}{x(t)-q(\epsilon t)},
\end{align}
we can reduce Eqs.~(\ref{eq. msl1}) and (\ref{eq. msl2}) to the following generalized phase equation:
\begin{align}
 \frac{d\theta(t)}{dt} &= e^{2q(\epsilon t)} - \epsilon e^{q(\epsilon t)} \sin\theta \cdot \dot{q}(\epsilon t)
 + \sigma [e^{2 q(\epsilon t)} - e^{q(\epsilon t)}\cos\theta] \cdot p(t).
\end{align}

We generated the input to the oscillator $I(t)$ by a Fourier series given by
\begin{align}
 I(t) &= \sum_{\ell=1}^{500} A_\ell \sin(\alpha_\ell t + \beta_\ell),
\end{align}
where $\beta_\ell$ is an {\it i.i.d.} random number drawn from a uniform distribution in $[0,2\pi]$,
\begin{align}
 \quad A_{\ell} &= \sqrt{10^{-3}\frac{\gamma}{1+\gamma^2\alpha_\ell^2}}, \\
 \alpha_\ell &= 0.1(\ell - 0.5),
\end{align}
and $\gamma$ is a parameter representing the characteristic scale of $I(t)$.
In this case, the long-time average of $I(t)$ becomes zero.
Thus, the criterion~(\ref{eq. formula}) gives a threshold frequency
\begin{align}
 \lambda_c &= \lambda(q)\big|_{q=0} = 2.
\end{align}

We computed the time series of the state variable $\bm{X}(t)=[x(t),y(t)]^{\top}$ through direct numerical simulations of Eqs.~(\ref{eq. msl1}) and (\ref{eq. msl2}),
and evaluated the approximation accuracy of the generalized phase equation (\ref{eq. phase eq}) by the mean square error ${\rm MSE}(\Omega_d)$ defined as
\begin{align}
 {\rm MSE}(\Omega_d) &= \frac{1}{\tau} \int_0^{\tau} \left[ \dot{\theta}_{\rm exact}(t) - \dot{\theta}_{\rm approx}(t) \right]^2 dt, \label{eq. mse exper}
\end{align}
where $\dot{\theta}_{\rm approx}(t)$ is a predicted value of $\dot{\theta}(t)$ obtained by plugging $\theta(t) = \Theta(x(t),y(t),q(\epsilon t))$
into the generalized phase equation (\ref{eq. phase eq}),
and $\dot{\theta}_{\rm exact}(t)$ is the exact value of $\dot{\theta}(t)$ given by
\begin{align}
 \dot{\theta}_{\rm exact}(t) &= \frac{\partial \Theta(x(t),y(t),q(\epsilon t))}{\partial x} F_1(x(t),y(t),I(t))
 + \frac{\partial \Theta(x(t),y(t),q(\epsilon t))}{\partial y} F_2(x(t),y(t),I(t)) \cr
 & \quad + \frac{\partial \Theta(x(t),y(t),q(\epsilon t))}{\partial q} \frac{dq(\epsilon t)}{dt}, \label{eq. exa val}
\end{align}
which can be directly calculated from the time series of $\bm{X}(t)$, $q(\epsilon t)$ and $I(t)$.
\INS{Note that the exact value of $\dot{\theta}(t)$ (Eq.~(\ref{eq. exa val})) also depends on the value of $\Omega_d$, because the definition of the phase variable $\theta(t)$ (Eq.~(\ref{eq. mse exper})) itself depends on $\Omega_d$.}

Figure~\ref{fig. 1} shows the time series of $I(t)$, $q(\epsilon t)$ and $\sigma p(t)$ for $\Omega_d=0.5,2,5$ and $10$ and $\gamma=5$.
Figure~\ref{fig. 2} shows the mean square error ${\rm MSE}(\Omega_d)$.
We see that ${\rm MSE}(\Omega_d)$ has a minimum for each value of the parameter $\gamma$,
which is in reasonable agreement with the criterion~(\ref{eq. formula}), {\it i.e.}, $\lambda_c=2$.
Though the criterion~(\ref{eq. formula}) proposed in this paper is based on a simplifying assumptions,
these results indicate that our criterion is able to give a reasonable threshold frequency $\Omega_d$.

\section{Conclusion}

In this paper, we proposed a simple criterion of the threshold frequency  for timescale decomposition of the external input for generalized phase reduction of limit-cycle oscillators.
Under the assumption\INS{s} that the amplitude relaxation is sufficiently fast and the \INS{timescale of the amplitude relaxation can be characterized by the second largest Floquet exponent of the oscillator when it is driven by a constant long-time average of the input,
we derived a criterion for choosing the threshold frequency, which is simple and physically reasonable}.
We confirmed the validity of our criterion by direct numerical simulations.
The criterion proposed in this paper is simple and easy to use, and thus it will be helpful in the engineering applications of the generalized phase reduction method, {\it e.g.},
optimal design of circuit oscillators~\cite{Vytyaz2009,Maffezzoni2010a}
and optimal control of periodically spiking neurons~\cite{Dasanayake2011,Nabi2012}.

\section*{Acknowledgments}

Financial support by KAKENHI (25540108, 26103510, 26120513) and CREST Kokubu project of JST are gratefully acknowledged.
One of the authors (WK) is supported by Grant-in-Aid for JSPS Fellows.

\section*{Appendix A: Variable transformation}

\INS{In the following appendices, we briefly review the derivation of the generalized phase equation~(\ref{eq. phase eq}), including the transformation of the state variable $\bm{X}(t)$ to the phase variable $\theta(t)$ and the amplitude variable $r(t)$, and the relation between the sensitivity functions, Eq.~(\ref{eq. der1}).  The results shown here are the same as those given in the Supplementary Information of our previous paper, Ref.~\cite{Kurebayashi2013}.

We consider a limit-cycle oscillator whose dynamics depends on a time-varying input $\bm{I}(t)$:
\begin{eqnarray}
 \dot{\bm{X}}(t)=\bm{F}(\bm{X}(t),\bm{I}(t)).
	\label{eq. model0}
\end{eqnarray}
The state variable $\bm{X}(t)$ is assumed to be two-dimensional here, but the result can be extended to higher-dimensional cases.
As argued in the Supplementary Information of Ref.~\cite{Goldobin} by Goldobin et al.,
we can define a phase $\theta=\Theta(\bm{X},\bm{I})$
and an amplitude $r=R(\bm{X},\bm{I})$ of the oscillator satisfying
\begin{eqnarray}
 \frac{\partial\Theta(\bm{X},\bm{I})}{\partial\bm{X}}\cdot
\bm{F}(\bm{X},\bm{I})&=&\omega(\bm{I}), \label{eq. theta def}\\
 \frac{\partial R(\bm{X},\bm{I})}{\partial\bm{X}}\cdot
\bm{F}(\bm{X},\bm{I})&=&-\lambda(\bm{I})R(\bm{X},\bm{I}), \label{eq. r def}
\end{eqnarray}
where $\lambda(\bm{I})$ is the absolute value of the second Floquet exponent
of the oscillator for constant $\bm{I}$.
Thus,
\begin{align}
\dot{\theta}(t)=\omega(\bm{I}),
\quad
\dot{r}(t)=-\lambda(\bm{I})r
\end{align}
when the input $\bm{I}$ is constant and in the given range $A \subset \mathbb{R}^m$.

When the parameter $\bm{I}(t)$ varies with time, we decompose $\bm{I}(t)$ into a slowly varying component $\bm{q}(\epsilon t)$ and remaining weak fluctuations $\sigma\bm{p}(t)$ as $\bm{I}(t) = {\bm q}(\epsilon t) + \sigma \bm{p}(t)$, and define the phase $\theta(t)$ and the amplitude $r(t)$ of the oscillator as
\begin{eqnarray}
 \theta(t)&=&\Theta(\bm{X}(t),\bm{q}(\epsilon t)),\\
 r(t)&=&R(\bm{X}(t),\bm{q}(\epsilon t)).
\end{eqnarray}
The dynamical equations for $\theta(t)$ and $r(t)$ are then given by
\begin{eqnarray}
 \dot{\theta}&=& 
	\left.\frac{\partial\Theta(\bm{X},\bm{I})}{\partial\bm{X}}
	\right|_{(\bm{X},\bm{q}(\epsilon t))}
	\cdot\frac{d\bm{X}(t)}{dt}
	+\left.\frac{\partial\Theta(\bm{X},\bm{I})}{\partial \bm{I}}
	\right|_{(\bm{X},\bm{q}(\epsilon t))}
	\cdot\frac{d\bm{q}(\epsilon t)}{dt}, \label{eq. theta2}\\
 \dot{r}&=&
	\left.\frac{\partial R(\bm{X},\bm{I})}{\partial\bm{X}}
	\right|_{(\bm{X},\bm{q}(\epsilon t))}
  \cdot\frac{d\bm{X}(t)}{dt}
	+\left.\frac{\partial R(\bm{X},\bm{I})}{\partial \bm{I}}
	\right|_{(\bm{X},\bm{q}(\epsilon t))}
	\cdot\frac{d\bm{q}(\epsilon t)}{dt}. \label{eq. r2}
	\label{eq. r exact2}
\end{eqnarray}
Plugging $\bm{I}(t)=\bm{q}(\epsilon t)+\sigma\bm{p}(t)$
into Eq.~(\ref{eq. model0}) and expanding it in
$\sigma$, we obtain
\begin{eqnarray}
 \dot{\bm{X}}=\bm{F}(\bm{X},\bm{q}(\epsilon t))
	+\sigma\bm{G}(\bm{X},\bm{q}(\epsilon t))\bm{p}(t)+O(\sigma^2),
	\label{eq. expand}
\end{eqnarray}
where $\bm{G}(\bm{X},\bm{q})$ is the matrix defined in the main article
as $G^{(j,k)}(\bm{X},\bm{q}) := \frac{\partial F_j(\bm{X},\bm{q})}{\partial q_k}$ ($j, k = 1, 2$ here).
Substituting Eqs.~(\ref{eq. theta def}), (\ref{eq. r def}),
and (\ref{eq. expand}) into Eqs.~(\ref{eq. theta2}) and (\ref{eq. r2}), we obtain
\begin{align}
 \dot{\theta}&=
	\omega(\bm{q}(\epsilon t))
	+\sigma\left.\frac{\partial\Theta(\bm{X},\bm{I})}{\partial\bm{X}}
	\right|_{(\bm{X},\bm{q}(\epsilon t))}
	\cdot\bm{G}(\bm{X},\bm{q}(\epsilon t))\bm{p}(t)
	+\epsilon\left.\frac{\partial\Theta(\bm{X},\bm{I})}{\partial \bm{I}}
	\right|_{(\bm{X},\bm{q}(\epsilon t))}
	\cdot\dot{\bm{q}}(\epsilon t)+O(\sigma^2), \label{eq. theta3} \\
 \dot{r}&=
	-\lambda(\bm{q}(\epsilon t))r
	+\sigma\left.\frac{\partial R(\bm{X},\bm{I})}{\partial\bm{X}}
	\right|_{(\bm{X},\bm{q}(\epsilon t))}
  \cdot\bm{G}(\bm{X},\bm{q}(\epsilon t))\bm{p}(t)
 +\epsilon\left.\frac{\partial R(\bm{X},\bm{I})}{\partial \bm{I}}
	\right|_{(\bm{X},\bm{q}(\epsilon t))}
	\cdot\dot{\bm{q}}(\epsilon t)+O(\sigma^2), \label{eq. r3}
\end{align}
where $\dot{\bm{q}}(\epsilon t) = d\bm{q}(\epsilon t)/d(\epsilon t)$.
By defining
$\bm{\zeta}_{\theta}(\theta,r,\bm{I})\in\mathbb{R}^m$,
$\bm{\zeta}_r(\theta,r,\bm{I})\in\mathbb{R}^m$,
$\bm{\xi}_{\theta}(\theta,r,\bm{I})\in\mathbb{R}^m$ and
$\bm{\xi}_r(\theta,r,\bm{I})\in\mathbb{R}^m$, respectively, as
\begin{eqnarray}
\bm{\zeta}_{\theta}(\theta,r,\bm{I})&=&
 \left.\bm{G}(\bm{X},\bm{I})^{\top}
	\frac{\partial\Theta(\bm{X},\bm{I})}{\partial\bm{X}}
	\right|_{\bm{X}=\bm{X}(\theta,r,\bm{I})}, \label{eq. zeta1}\\
 \bm{\zeta}_r(\theta,r,\bm{I})&=&
	\left.\bm{G}(\bm{X},\bm{I})^{\top}
 \frac{\partial R(\bm{X},\bm{I})}{\partial\bm{X}}
 \right|_{\bm{X}=\bm{X}(\theta,r,\bm{I})}, \label{eq. zeta2} \\
 \bm{\xi}_{\theta}(\theta,r,\bm{I})&=&
	\left.\frac{\partial\Theta(\bm{X},\bm{I})}{\partial\bm{I}}
	 \right|_{\bm{X}=\bm{X}(\theta,r,\bm{I})}, \label{eq. xi1} \\
 \bm{\xi}_r(\theta,r,\bm{I})&=&
 \left.\frac{\partial R(\bm{X},\bm{I})}{\partial\bm{I}}
				\right|_{\bm{X}=\bm{X}(\theta,r,\bm{I})}, \label{eq. xi2}
\end{eqnarray}
where $\bm{X}(\theta,r,\bm{I})$
$\in \mathbb{R}^2$ represents an oscillator state with 
$\theta=\Theta(\bm{X},\bm{I})$, $r=R(\bm{X},\bm{I})$,
and parameter ${\bm I}$,
Eqs.~(\ref{eq. theta3}) and (\ref{eq. r3}) can be written as
\begin{eqnarray}
  \dot{\theta}&=&
	\omega(\bm{q}(\epsilon t))
	+\sigma\bm{\zeta}_{\theta}(\theta,r,\bm{q}(\epsilon t))
	\cdot\bm{p}(t)
	+\epsilon\bm{\xi}_{\theta}(\theta,r,\bm{q}(\epsilon t))
	\cdot\dot{\bm{q}}(\epsilon t)+O(\sigma^2), \label{eq. theta4} \\
 \dot{r}&=&
-\lambda(\bm{q}(\epsilon t))r
+\sigma\bm{\zeta}_r(\theta,r,\bm{q}(\epsilon t))
  \cdot\bm{p}(t)
	+\epsilon\bm{\xi}_r(\theta,r,\bm{q}(\epsilon t))
	\cdot\dot{\bm{q}}(\epsilon t)+O(\sigma^2). \label{eq. r4}
\end{eqnarray}
Here, $\bm{\zeta}_{\theta}(\theta,0,\bm{I})$ and $\bm{\xi}_{\theta}(\theta,0,\bm{I})$ with $r=0$ correspond to the sensitivity functions $\bm{\zeta}(\theta,\bm{I})$ and $\bm{\xi}(\theta,\bm{I})$ defined in the main article.
The other two functions $\bm{\zeta}_r(\theta,r,\bm{I})$ and $\bm{\xi}_r(\theta,r,\bm{I})$ represent sensitivities of the amplitude variable to the small fluctuations and to the slowly varying component of the input, respectively.

When $\lambda({\bm q}(\epsilon t))$ is sufficiently large, the amplitude $r(t)$ takes tiny values around $0$, and the phase $\theta(t)$ approximately obeys Eq.~(\ref{eq. theta4}) with $r=0$, i.e., the
generalized phase equation~(\ref{eq. phase eq}), which yields an approximation error of $O(r)$.  See Ref.~\cite{Kurebayashi2013} for a detailed discussion on the validity of the approximation.

\section*{Appendix B: Derivation of Eq.~(\ref{eq. der1})}

In this Appendix, we derive Eq.~(\ref{eq. der1}). As explained in the main article, we assume the existence of a stable limit-cycle orbit ${\bm X}_{0}(t, {\bm I})$ with frequency $\omega({\bm I})$ satisfying $d{\bm X}_{0}(t, {\bm I}) / dt = {\bm F}({\bm X}_{0}(t, {\bm I}), {\bm I})$ that smoothly depends on ${\bm I}$ for each constant ${\bm I} \in A$.
We denote the phase of the oscillator as $\theta(t) = \omega({\bm I}) t$, and represent the oscillator state as ${\bm X}_{0}(t, {\bm I}) = {\bm X}_{0}(\theta(t), {\bm I})$ by using the phase $\theta(t)$ instead of $t$.

We differentiate both sides of Eq.~(\ref{eq. r def}) with respect to $\bm{I}$ and plug in $\bm{X}=\bm{X}_0(\theta, \bm{I})$. From the left-hand side, we obtain
\begin{align}
 \left. \frac{\partial}{\partial\bm{I}} \left[ \frac{\partial R(\bm{X},\bm{I})}{\partial\bm{X}} \cdot \bm{F}(\bm{X},\bm{I}) \right] \right|_{\bm{X} = \bm{X}_0(\theta,\bm{I})}
 &= \left.\left[\frac{\partial}{\partial\bm{X}}\left(\frac{\partial R(\bm{X},\bm{I})}{\partial\bm{I}}\right)\right]
 \bm{F}(\bm{X},\bm{I})\right|_{\bm{X}=\bm{X}_0(\theta,\bm{I})}
 + \bm{\zeta}_r(\theta,0,\bm{I}),
 \label{eq. def R func def}
\end{align}
where we used the definition of ${\bm G}({\bm X}, {\bm I})$ and Eq.~(\ref{eq. zeta2}) with $r=0$.
The first term on the right-hand side can be further calculated as
\begin{align}
 &\left.\left[\frac{\partial}{\partial\bm{X}}\left(\frac{\partial R(\bm{X},\bm{I})}{\partial\bm{I}}\right)\right]
 \bm{F}(\bm{X},\bm{I})\right|_{\bm{X}=\bm{X}_0(\theta,\bm{I})}
 =\left.\left[\frac{\partial}{\partial\bm{X}}\left(\frac{\partial R(\bm{X},\bm{I})}{\partial\bm{I}}\right)\right]
 \right|_{\bm{X}=\bm{X}_0(\theta,\bm{I})}
 %%%\left.\frac{d\bm{X}_0(\theta(t),\bm{I})}{dt}\right|_{t=\theta/\omega(\bm{I})} \cr
\frac{d\bm{X}_0(\theta(t),\bm{I})}{dt} \cr
 &= \omega(\bm{I}) \left.\left[\frac{\partial}{\partial\bm{X}}\left(\frac{\partial R(\bm{X},\bm{I})}{\partial\bm{I}}\right)\right]
 \right|_{\bm{X}=\bm{X}_0(\theta,\bm{I})}
 \frac{\partial\bm{X}_0(\theta,\bm{I})}{\partial\theta}
 = \omega(\bm{I}) \frac{\partial}{\partial\theta} \left. \left( \frac{\partial R(\bm{X},\bm{I})}{\partial \bm{I}} \right)
 \right|_{\bm{X}=\bm{X}_0(\theta,\bm{I})} \cr
 &= \omega(\bm{I}) \frac{\partial \bm{\xi}_r(\theta,0,\bm{I})}{\partial \theta},
 \label{eq. def R func def - rewrite}
\end{align}
where we used the chain rule for the derivative $\partial / \partial \theta$
and Eq.~(\ref{eq. xi2}).
Similarly, by differentiating the right-hand side of Eq.~(\ref{eq. r def}) with respect to ${\bm I}$, we can derive
\begin{align}
 \left. \frac{\partial}{\partial\bm{I}} \left[ - \lambda(\bm{I}) R(\bm{X},\bm{I}) \right] \right|_{\bm{X}=\bm{X}_0(\theta,\bm{\bm{I}})}
 &= - \left. \left[ \frac{d\lambda(\bm{I})}{d\bm{I}} R(\bm{X},\bm{I})
 + \lambda(\bm{I}) \frac{\partial R(\bm{X},\bm{I})}{\partial\bm{I}} \right] \right|_{\bm{X}=\bm{X}_0(\theta,\bm{\bm{I}})} 
 = -  \lambda(\bm{I}) \bm{\xi}_r(\theta,0,\bm{I}),
 \label{eq. def R func def left}
\end{align}
where we used $R(\bm{X},\bm{I})\big|_{\bm{X}=\bm{X}_0(\theta,\bm{I})} = 0$.
Thus, we obtain a linear first-order ordinary differential equation for $\bm{\xi}_r(\theta,0,\bm{I})$,
\begin{align}
 \omega(\bm{I}) \frac{\partial \bm{\xi}_r(\theta,0,\bm{I})}{\partial \theta} + \bm{\zeta}_r(\theta,0,\bm{I})
 &= -  \lambda(\bm{I}) \bm{\xi}_r(\theta,0,\bm{I}),
 \label{eq. R 1st ode}
\end{align}
which can be solved as
\begin{align}
 \bm{\xi}_r(\theta,0,\bm{I})
 &= - \frac{1}{\omega(\bm{I})} \int_{-\infty}^{\theta} e^{\lambda(\bm{I})(\theta'-\theta)/\omega(\bm{I})} \bm{\zeta}_r(\theta',0,\bm{I}) d\theta'
=
\frac{1}{\lambda({\bm I})} \int_{0}^{\infty} e^{-s} \bm{\zeta}_r\left( \theta - \frac{\omega(\bm{I})}{\lambda(\bm{I})} s, 0, {\bm I}\right) ds.
\end{align}
By expanding the integrand, the order of ${\bm \xi}_{r}(\theta, 0, {\bm I})$ can be estimated as
\begin{align}
 \bm{\xi}_r(\theta,0,\bm{I})
 &= \frac{1}{\lambda(\bm{I})} \int_{0}^{\infty} e^{-s}  \bm{\zeta}_r ( \theta, 0, {\bm I} ) ds 
 + O\left( \frac{1}{\lambda({\bm I})^{2}} \right) \cr
 &= \frac{1}{\lambda(\bm{I})} \bm{\zeta}_r(\theta, 0, \bm{I})
 + O\left(\frac{1}{\lambda(\bm{I})^2}\right),
 \label{eq. order xi r}
\end{align}
which gives Eq.~(\ref{eq. der1}).
}


\begin{thebibliography}{99}

\bibitem{kuramoto}
Y. Kuramoto, {\it {Chemical Oscillations, Waves, and Turbulence}},
Springer, Berlin, 1984.

\bibitem{winfree}
A. T. Winfree, {\it {The Geometry of Biological Time}},
Springer, Berlin, 1980.

\bibitem{pikovsky}
A. Pikovsky, M. Rosenblum, and J. Kurths, {\it {A universal concept in nonlinear sciences}},
Cambridge Univ. Press, Cambridge, 2001.

\INS{
\bibitem{Demir2000}
A. Demir, A. Mehrotra, and J. Roychowdhury,
\newblock {\it IEEE Trans. Circuits Syst.-I. Fundam. Theory Applicat.} 47, 655--674, 2000.

\bibitem{Tanaka2002}
H.-A. Tanaka, A. Hasegawa, H. Mizuno, and T. Endoh,
\newblock {\it IEEE Trans. Circuits Syst.-I. Fundam. Theory Applicat.} 49, 1271--1278, 2002.

\bibitem{Nagashima2014}
T. Nagashima, X. Wei, H. A. Tanaka, and H. Sekiya,
\newblock {\it IEEE Trans. Circuits Syst.-I. Fundam. Theory Applicat.} 61, 2904-2911, 2014.
}

\bibitem{Vytyaz2009}
I. Vytyaz, D. C Lee, and P. K. Hanumolu,
\newblock {\it IEEE Trans. Circuits Syst.-I. Fundam. Theory Applicat.} 28, 609--622, 2009.

\bibitem{Maffezzoni2010a}
P. Maffezzoni, D. D. Amore, S. Daneshgar, and M. P. Kennedy,
\newblock {\it IEEE Trans. Circuits Syst.-I. Fundam. Theory Applicat.} 57, 2956--2966, 2010.

\INS{
\bibitem{Kure2014}
W. Kurebayashi, T. Ishii, M. Hasegawa and H. Nakao,
\newblock {\it Europhys. Lett.} 107, 10009, 2014.
}

\bibitem{Dasanayake2011}
I. Dasanayake and J.-S. Li,
\newblock {\it Phys. Rev. E} 83, 061916, 2011.

\bibitem{Nabi2012}
A. Nabi and J. Moehlis,
\newblock {\it J. Math. Biol.} 64, 981--1004, 2012.

\INS{
\bibitem{Moehlis2014}
G. S. Schmidt, D. Wilson, F. Allg\"ower, and J. Moehlis
\newblock {\it Nonlinear Theory and Its Applications} 5, 424--435, 2014.
}

\bibitem{Kurebayashi2013}
W. Kurebayashi, S. Shirasaka, and H. Nakao,
\newblock {\it Phys. Rev. Lett.} 111, 214101, 2013.

\bibitem{Goldobin}
D. S. Goldobin,  J. Teramae, H. Nakao, and G. B. Ermentrout,
Phys. Rev. Lett. {\bf 105}, 154101 (2010).

\if 0
\bibitem{COWT}
Y. Kuramoto, 
{\it Chemical Oscillations, Waves and Turbulence}
(Dover, New York, 2003).

\bibitem{SUCNS} 
A. Pikovsky, M. Rosenblum, and J. Kurths,
{\it Synchronization: A Universal Concept in Nonlinear Sciences}
(Cambridge University Press, Cambridge, 2001).

\bibitem{MFN}
G. B. Ermentrout and D. H. Terman, {\it Mathematical Foundations of Neuroscience} (Springer, New York, 2010).
\fi

\end{thebibliography}
\end{document}